 \documentclass[preprint2]{aastex}

\slugcomment{To appear in AJ}

\shorttitle{The evolution of radio galaxy hosts}
\begin{document}


\title{The evolution of the stellar hosts of radio galaxies}


\author{Mark Lacy}
\affil{Institute of Geophysics and Planetary Physics, L-413 Lawrence Livermore
National Laboratory, Livermore CA~94550, and Dept.\ of Physics, 
University of California, 1 Shields Avenue, Davis CA~95616 }
\email{mlacy@igpp.ucllnl.org}

\author{Andrew J.\ Bunker \altaffilmark{1}}
\affil{Department of Astronomy, University of California at Berkeley, \\
601 Campbell Hall, Berkeley CA~94720, USA}
\email{bunker@ast.cam.ac.uk}

\and

\author{Susan E.\ Ridgway}
\affil{Bloomberg Center for Physics and Astronomy, Johns Hopkins University, 
3400 N.\ Charles Street, Baltimore, MD21218}
\email{ridgway@pha.jhu.edu}

\altaffiltext{1}{Current address: Institute of Astronomy, Madingley Road, 
Cambridge, CB3 OHA, UK}

\begin{abstract}
We present new near-infrared images of $z>0.8$ radio galaxies from the 
flux-limited 7C-{\sc iii} sample of radio sources for which we have recently 
obtained almost complete spectroscopic redshifts. The 7C objects 
have radio luminosities $\approx 20$ times fainter than 3C radio galaxies at 
a given redshift. The absolute magnitudes of the underlying host
galaxies and their scale sizes are only weakly dependent on radio 
luminosity. Radio galaxy hosts at $z\sim 2$ are significantly brighter 
than the hosts of radio-quiet quasars at similar redshifts and the
model AGN hosts of Kauffmann \& Haehnelt (2000). There is no evidence 
for strong evolution in scale size, which shows a large scatter at 
all redshifts. The hosts brighten significantly
with redshift, consistent with the passive evolution of a stellar population
that formed at $z\stackrel{>}{_{\sim}} 3$. This scenario is consistent with 
studies of host galaxy morphology and submillimeter continuum emission, both 
of which show strong evolution at $z\stackrel{>}{_{\sim}}2.5$. The lack of a 
strong ``redshift cutoff'' in the radio luminosity function to $z>4$ suggests 
that the formation epoch of the radio galaxy host population 
lasts $\stackrel{>}{_{\sim}}1$ Gyr from $z\stackrel{>}{_{\sim}}5$ to 
$z\sim 3$. We suggest these facts are best explained by models in which the 
most massive galaxies and their associated AGN form early due to high baryon 
densities in the centres of their dark matter haloes. 
\end{abstract}


\keywords{galaxies: evolution -- galaxies: active -- galaxies: formation}


\section{Introduction}

At low redshifts, FRI and FRII radio sources with radio luminosities 
at 151 MHz of $L_{R (151)} \stackrel{>}{_{\sim}} 10^{24} {\rm WHz^{-1}}$ are 
associated 
almost exclusively with giant elliptical host galaxies. If this continues to 
be the case out to high redshift, then radio galaxies can give us a unique 
insight into the formation and evolution of a single class of massive galaxy. 
This is particularly exciting in the light of submillimetre detections of 
$z\sim 4$ radio galaxies (e.g.\ Archibald et al.\ 2000) which may indicate 
that we can see these objects
during their major bursts of star formation [although see also Willott, 
Rawlings \& Jarvis (2000)]. Furthermore, the similarity of radio galaxy hosts, 
in contrast to the wide range in luminosity of radio-quiet quasar hosts
(McLure et al.\ 1999; Ridgway et al.\ 2000), suggests that one of the 
conditions necessary for producing powerful radio jets is the presence of a 
massive spheroidal component, and therefore a supermassive 
($\stackrel{>}{_{\sim}} 10^9 M_{\odot})$ black hole (Lacy, Ridgway \& 
Trentham 2000).

Studies of high redshift radio galaxy hosts have traditionally concentrated
on the $K-z$ Hubble Diagram. Work on the 3C and 1 Jy samples by 
Lilly (1989 and refs.\ therein) initially pointed to a passively-evolving 
stellar host formed at high redshift which evolved into the giant elliptical
radio galaxy hosts seen today, but this was challenged when some 
high redshift radio galaxies were found to have significant emission line
contributions to their $K$-band light (Eales \& Rawlings 1993, 1996). 
Only by finding low AGN-luminosity radio galaxies at high redshift could the 
controversy be
resolved. Eales et al.\  (1997) used the 6C sample, a factor of five fainter
in radio flux than the 3C sample, to show that there did seem to be a radio 
luminosity dependence of host galaxy magnitude. Further work by 
Roche, Eales \& Rawlings (1998) indicated that the hosts of 6C radio galaxies
were not only significantly fainter than their 3C counterparts, but also 
had smaller scale sizes.

We have 
used the 7C-{\sc iii} radio galaxy redshift survey of Lacy et al.\ (1999b) 
to select a complete sample of $z>0.8$ radio galaxies. The 7C redshift 
surveys are a factor of 4--5 lower still in luminosity at a given redshift 
than the 6C sample of Eales et al.\ (1997), and thus allow the study of 
high redshift radio galaxy hosts over a wide range in radio luminosity
(Willott et al.\ 1999). The 7C-{\sc iii} sources were imaged 
in the near-infrared on the 3-m NASA Infrared Telescope Facility
(IRTF) and the 3-m Shane Telescope at Lick Observatory. 
These data have allowed us to further investigate the radio
luminosity dependence of host properties at $z\sim 1$ and, because our 
7C objects at $z\sim 2$ 
have similar radio
luminosities to $z\sim 1$ 6C radio galaxies and $z\sim 0.3$ 3C 
radio galaxies, we can also investigate host galaxy evolution over a wide 
range in redshift. 

We assume a cosmology with $\Omega_{\rm M} =1, 
\Omega_{\Lambda}=0$ and $H_0=50 \, {\rm kms^{-1}Mpc^{-1}}$ except where 
otherwise stated.

\begin{deluxetable}{llclll}
\footnotesize
\tablecaption{Observing log \label{tbl-1}}
\tablewidth{0pt}
\tablehead{
\colhead{Object} & \colhead{Telescope} & \colhead{$z$} & 
\colhead{Filter} & \colhead{$t_{\rm int}$/min}   
& \colhead{FWHM PSF/$^{''}$}} 

\startdata
7C 1733+6719 &IRTF &1.84 & $K^{'}$ & 18  & 0.95\\
7C 1740+6640 &IRTF &2.10 & $K^{'}$ & 18  & 0.65\\
7C 1741+6704 &IRTF &1.05 & $J$     & 13.5& 0.65\\
7C 1742+6346 &IRTF &1.27 & $H$     & 18  & 0.90\\
7C 1748+6703 &Shane& ?   & $J$ \& $K^{'}$& 56 & 1.3\\
             &     &     &         &     &     \\
7C 1748+6657 &IRTF &1.05 & $J$     & 13.5& 0.90\\
7C 1751+6809 &IRTF &1.54 & $H$     & 18  & 0.56\\
7C 1753+6311 &Shane&1.96?& $J$ \& $K^{'}$& 82 & 1.5\\
7C 1754+6420 &IRTF &1.09 & $J$     & 13.5& 1.07\\
7C 1756+6520 &IRTF &1.48?& $K^{'}$ & 18  & 0.78\\
             &     &     &         &     &     \\
7C 1758+6719 &IRTF &2.70 & $K^{'}$ & 82  & 0.42\\
7C 1802+6456 &IRTF &2.11 & $K^{'}$ & 18  & 0.74\\
7C 1804+6313 &IRTF &?    & $K^{'}$ & 18  & 0.80\\  
7C 1805+6332 &IRTF &1.84 & $K^{'}$ & 36  & 0.63\\
7C 1807+6841 &IRTF &0.82 & $J$     & 13.5& 0.72\\
             &     &     &         &     &     \\
7C 1807+6719 &IRTF &2.78 & $K^{'}$ & 36  & 0.75\\
7C 1812+6814 &IRTF &1.08 & $J$     & 13.5& 0.93\\
7C 1814+6702 &IRTF &4.05?& $K^{'}$ & 54  & 0.78\\
             &Shane&     & $K^{'}$ & 54  & 1.3\\
7C 1814+6529 &IRTF &0.96 & $J$     & 13.5& 0.84\\
7C 1820+6657 &IRTF &2.98 & $K^{'}$ & 36  & 0.69\\
             &     &     &         &     &     \\
7C 1816+6605 &IRTF &0.92 & $J$     & 13.5& 1.02\\
7C 1825+6602 &IRTF &2.38 & $K^{'}$ & 18  & 0.81\\
 \enddata
\end{deluxetable}

\section{Observations}

Most sources were observed on the IRTF with NSFCAM, a near infrared imaging 
camera employing a $256\times 256$ InSb array, on the nights of 1999 July 
28 -- 29 UT. The 0.3 arcsec/pixel scale was used for all observations. Details
of the observations are given in Table 1. One of the $J$, $H$ or $K^{'}$ 
filters was chosen for each object
so as to cover as far as possible the rest-frame $R$-band, thus 
minimizing the $k$-correction and the associated uncertainties. In practice
this meant that objects with $0.8<z<1.2$ were observed in $J$-band, those 
with $1.2<z<1.8$ in $H$-band and those with $z>1.8$ in $K$-band 
(using a $K^{'}$ filter). Exceptions to 
this were two objects with uncertain redshifts, 7C 1756+6520 and 
7C 1804+6313, which were both observed in $K^{'}$ for ease of estimating
photometric redshifts. The sky was clear throughout the run, and most 
objects were observed in conditions of sub-arcsecond seeing. 
Two more objects with uncertain redshifts were observed using the Gemini 
instrument (McLean et al.\ 1993, 1994) on the Shane 
Telescope at Lick Observatory on the nights of 1999 June 1 and 4 UT. 
Gemini was used with a dichroic beamsplitter 
which enabled us to image in $J$ and $K^{'}$ simultaneously. The detector on 
the short wavelength arm was a $256\times 256$ HgCdTe array, and that 
on the long wavelength arm a $256\times 256$ InSb array. The image
scale in both arms was 0.68 arcsec/pixel.
7C~1814+6704 was observed in both runs, but the better seeing of the IRTF
data allowed more accurate photometry in the moderately crowded field so 
only the IRTF data is presented. In addition 
a $K^{'}$ image of 7C~1745+6624 was presented in Lacy et al.\ (1999b).

The data were reduced using the {\sc dimsum} package in {\sc iraf}, and 
flat-fielded using dome flats. The final images were magnified by a 
factor of two before combination to improve the sampling of the final image.
The reduced images are shown in Fig.\ 1 and the photometric properties detailed
in Table 2. 

\onecolumn
\begin{figure}
\begin{picture}(500,550)
\put(0,-100){\includegraphics[scale=0.85]{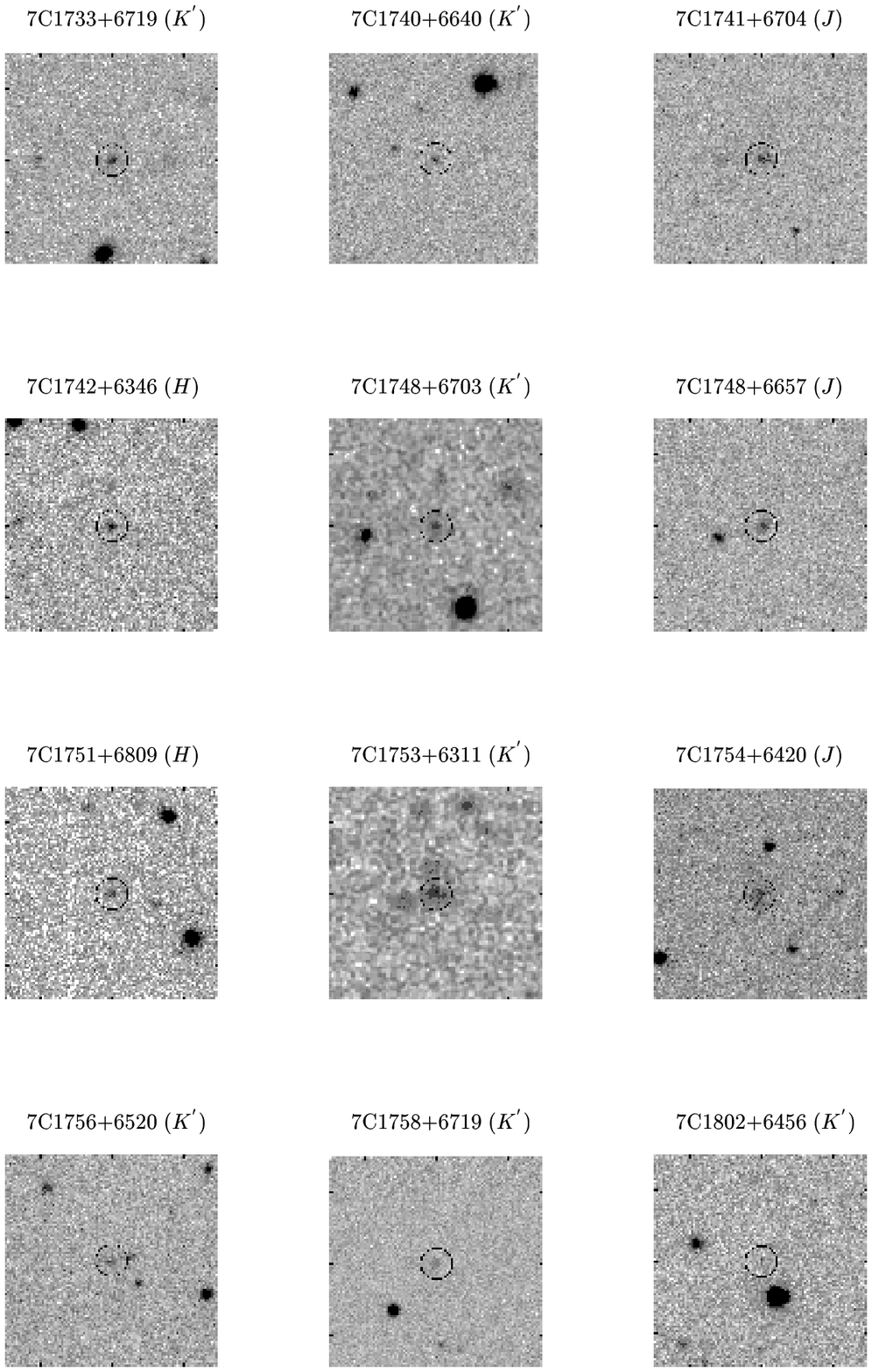}}
\end{picture}
\caption{Near-infrared images of the $z>0.8$ 7C sources. The
position of each identification is indicated by the dotted circle. The images
are $\approx 30$ arcsec on a side.}
\end{figure}
\setcounter{figure}{0}
\begin{figure}
\begin{picture}(500,550)
\put(0,-100){\includegraphics[scale=0.85]{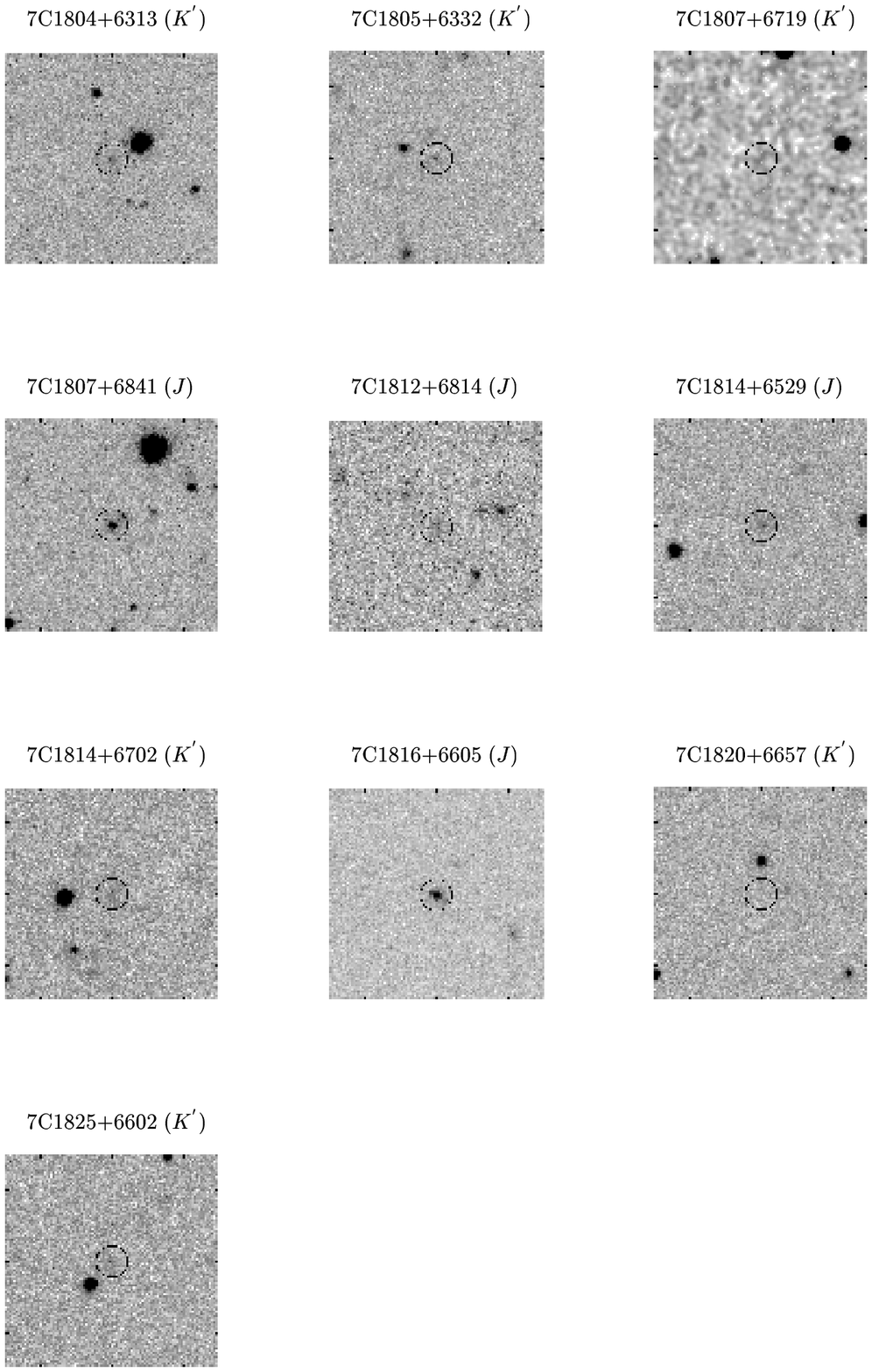}}
\end{picture}
\caption{continued}
\end{figure}

%\figcaption[fig1_pg2.ps]{Near-infrared images of the $z>0.8$ 7C sources. The
%position of each identification is indicated by the circle.}
\twocolumn

\section{Analysis}

\subsection{Magnitudes and $k$-corrections}

Aperture magnitudes were measured as listed in Table 2 and corrected to a 
standard 63.9 kpc metric aperture according to the prescription of Eales 
et al.\ (1997), which assumes a power-law curve of growth of the form 
$I(<r) \propto r^{0.35}$ for $z>0.6$ objects. Our metric aperture corresponds 
to about 8 arcsec at $z\stackrel{>}{_{\sim}} 1$ in our assumed cosmology, and 
this angular size is not strongly dependent on the choice of 
$\Omega_{\rm M}$ and $\Omega_{\Lambda}$. For many of our objects we have 
been able to measure the 
magnitude in a large aperture, but for some we have had to restrict the 
aperture to $\approx 3$ arcsec to avoid contamination from neighbouring 
objects or 
excessive noise in the case of faint objects. The aperture correction applied 
in these cases is $\approx 0.4$ magnitudes. The disadvantage of this 
technique is that the aperture magnitude and in particular the correction are 
scale-size dependent. An alternative would have been to use
total magnitudes (cf.\ Roche et al.\ 1998). 
However, we decided against this as for low 
signal-to-noise detections the curve of growth is ill-defined and can 
lead to large errors in the magnitudes (Stevens 1999). The mean half-light 
radius of our objects is 0.9 arcsec, so the typical amount 
of flux missed in an 8 arcsec aperture will be very small in any case. We thus 
believe that the use of these large aperture magnitudes will not seriously 
affect the results of this paper. 

For objects with spectroscopic redshifts, 
$k$-corrections to rest-frame $R$-band were made using 1Gyr-burst models
generated using the {\sc pegase} code (Fioc \& Rocca-Volmerange 1997), 
integrating over the appropriate filter profiles. A model
galaxy with a total age of 2 Gyr was used for objects with $2<z\leq 3$, and 
3 Gyr for objects with $0.8<z\leq 2$. The $z>3$ objects used a 1.4 Gyr-old 
model galaxy. For most objects the $k$-corrections are small ($< 0.2$ 
magnitudes), the exceptions to this being the $z>3$ objects for which 
$K^{'}$ corresponds to rest-frame wavelengths significantly shorter than 
$R$-band (but still above 4000\AA ) and 7C 1756+6520 which was observed in 
$K^{'}$ to check its redshift photometrically. The biggest 
correction was $-$0.61 mag.\ for 7C~1814+6704 at a probable redshift of 4.05.

\begin{deluxetable}{lclcrcrr}
\footnotesize
\tablecaption{Photometry and scale sizes of the 7C-{\sc iii} radio galaxies
\label{tbl-2}}
\tablewidth{0pt}
\tablehead{
\colhead{Object} & \colhead{$\phi$/$^{''}$} &
\colhead{Mag.\ } & 
\colhead{$M_{R}$} & \colhead{$r_{\rm hl}/^{''}$} & 
\colhead{Range/$^{''}$} & \colhead{$r_{\rm hl}$/kpc}& \colhead{\% lines}
}
% A: (1,0,50) B:(0.3,0.7,50), MR is in 63.9 kpc ap.

\startdata
7C 1733+6719 & 5 &$K^{'}=18.3 \pm 0.2$ &-24.7&1.1 & 0.46-1.9 &9.2 &0\\
7C 1740+6640 & 3 &$K^{'}=18.9 \pm 0.2$ &-24.6&$<$0.37& -     &3.0 &17\\
7C 1741+6704 & 8 &$J=19.7\pm 0.2$     &-23.6&2.0& 0.60-6.2   &17  &9\\
7C 1742+6346 & 3 &$H=19.2\pm 0.2$     &-23.9&$<$0.5& -    &$<$8.6 &7\\
7C 1745+6624 & 2 &$K^{'}=20.8\pm 0.3$ &-23.8&  -  & -       & -   &10\\
             &   &             &     &    &          &            &\\
7C 1748+6703 & 8 &$J=21.6\pm 0.3$     & -   &  -  & -     & -     &?\\
             &   &$K^{'}=18.3\pm 0.2$ & - & - & -         &-      &?\\
7C 1748+6657 & 8 &$J=19.6\pm 0.2$     &-23.8&0.90& 0.58-1.4 &7.7  &8\\
7C 1751+6809 & 3 &$H=19.5\pm 0.2$     &-23.9&0.22& $<$1.5   &1.9  &6\\
7C 1753+6311 & 3 &$J=20.7\pm 0.3$     & - & - & -        &-       &?\\
             &   & $K^{'}=18.6\pm 0.2$& - & - & -        &-       &?\\
7C 1754+6420 & 8 &$J=19.1\pm 0.2$     &-24.1&4.5 &1.6-13    &39   &10\\
             &   &             &     &    &          &            &\\
7C 1756+6520 & 3 &$K^{'}=19.2\pm 0.2$ & -   & -  &  -       & -   &0\\
7C 1758+6719 & 8 &$K^{'}=19.3\pm 0.2$ &-24.5&0.35&0.12-1.0  &2.6  &0\\
7C 1802+6456 & 8 &$K^{'}=19.3\pm 0.3$ &-24.1&0.97&0.15-7.0  &7.9  &50\\
7C 1804+6313 & 3 &$K^{'}=19.1\pm 0.2$ & -   & -  &  -       & -   &?\\
7C 1805+6332 & 5 &$K^{'}=19.0\pm 0.2$ &-23.5&1.9 &0.52-7.1  &16   &0\\
             &   &             &     &    &          &            &\\
7C 1807+6719 & 5 &$K^{'}=20.2\pm 0.3$ &-23.9& -  &   -      & -   &0\\
7C 1807+6841 & 8 &$J=18.8\pm 0.2$     &-23.7&0.87&0.54-1.4  &7.2  &6\\
7C 1812+6814 & 5 &$J=20.4\pm 0.3$     &-22.9&1.1 &0.17-10   &9.4  &5\\
7C 1814+6702 & 5 &$K^{'}=19.4\pm 0.2$ &-25.7&  - & -        &-    &0\\   
7C 1814+6529 & 8 &$J=19.2\pm 0.15$     &-23.9&0.89&0.35-2.3  &7.6 &$<$3\\
             &   &             &     &    &          &            &\\
7C 1816+6605 & 5 &$J=18.6\pm 0.15$     &-24.4&0.79&0.56-1.1  &6.7 &1\\
7C 1816+6710 & 5 &$H=19.0\pm 0.3$     &     &    &          &     &0 \\
7C 1820+6657 & 3 &$K^{'}>20.7$ &$>$-23.7&   -&     -    &  -      &$>$5 \\
7C 1825+6602 & 5 &$K^{'}=19.0\pm 0.2$ &-24.7&0.59&0.56-0.63 &4.6  &10\\
\enddata

\tablecomments{Errors on the magnitudes are estimated assuming a 
photometric error of 0.1 magnitude combined with noise. 
$\phi$ is the aperture diameter and $M_R$ is 
measured in a 63.9 kpc metric aperture. The limit on the magnitude of 
7C~1820+6657 is a 3-$\sigma$ limit.}

\end{deluxetable}

\onecolumn
\begin{figure}
\begin{picture}(500,200)
\includegraphics{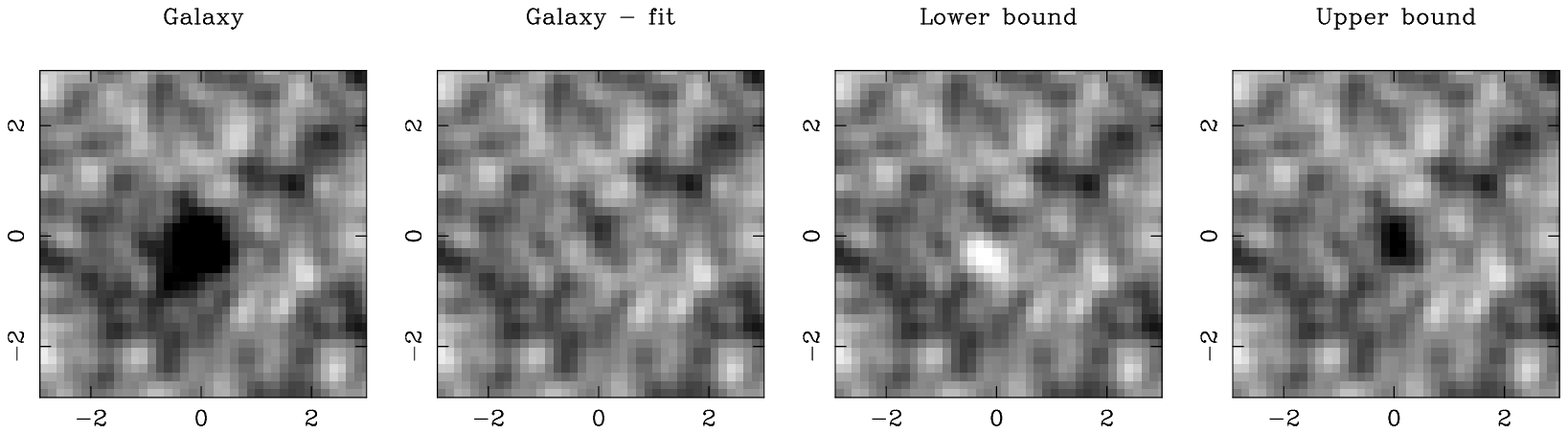}
\end{picture}
%\figcaption[figure2.ps]{An example of the model fitting procedure to the
\caption{An example of the model fitting procedure to the
galaxy 7C 1758+6719. The galaxy itself, the residual of the 
galaxy minus the best fit model ($r_{\rm hl}=0.35$ arcsec)
and the galaxy minus the approximate $\pm 1 \sigma$ bounds on
$r_{\rm hl}$ (0.12 and 1.0 arcsec) are shown, running from left to right. 
All images have the 
same symmetric greyscale stretch, with black corresponding to a surface
brightness of $K^{'}=20.8$mag arcsec$^{-2}$,
and the subtractions have been arranged such that each
residual image contains zero net flux. Thus a model with too small a scale 
size results in a compact negative residual near the centre of the image where 
oversubtraction has occurred, and a model with too large a scale size leaves 
a compact positive residual near the centre. The flux in the compact 
residuals left after subtraction of the $\pm 1 \sigma$ models, 
measured in a 2-arcsec diameter aperture, are $\pm 0.2$ times the total
galaxy flux in the same aperture.}
\end{figure}
\twocolumn

\subsection{Photometric redshifts for objects in the 7C-{\sc iii} sample}

Several of the objects observed had uncertain redshifts [grade $\gamma$ 
in Lacy et al.\ (1999b)] or no redshift information at all. Of the
objects without spectroscopic redshifts, 7C~1748+6703 
has a $K^{'}$ magnitude and colours consistent with $2.4<z<4$ whereas 
7C~1753+6311 and 7C~1804+6313 both have magnitudes and colours 
consistent with them having redshifts in the range $1.2<z<1.8$
(Willott et al.\ 2000). In addition 
Lacy et al.\ (1999b) estimated a photometric redshift of 1.7 for 7C 1743+6341. 
The identification of 7C~1753+6311 in Lacy et al.\ (1999b), which had a 
provisional redshift of 1.96, actually corresponds to a blue galaxy along the 
radio axis. Our Lick image showed that the true
identification is a very red ($R-K^{'} \stackrel{>}{_{\sim}} 5$) 
galaxy 4.7 arcsec away which is significantly closer to the likely radio 
central component. Nevertheless, the original identification may well be 
associated with the radio galaxy, and the magnitude and colours are consistent
with $z\stackrel{<}{_{\sim}} 2$, so we have not revised our redshift 
estimate. As we aligned the slit with the radio axis, though, we might have 
expected to see Ly$\alpha$ in the spectrum if the redshift were indeed 
$1.96$. Thus we note that the redshift may well be lower.
7C~1756+6520 had a provisional redshift of 1.48, and its $K^{'}$ magnitude is 
just consistent with this, although somewhat fainter than the mean $K-z$ 
relation at this redshift. 7C~1814+6702 with a provisional redshift of 4.05 
has a $K^{'}$ magnitude in both the IRTF and Lick data 
consistent with both $z\sim 4$ and $1.2<z<1.8$, although 
its diffuse and possibly aligned structure is typical of $z>3$ objects
(van Breugel et al.\ 1998) so has been 
kept at 4.05 pending a deeper spectrum. 7C 1816+6710 was observed in $H$-band
by Lacy et al.\ (1999a), its magnitude is consistent with its 
provisional redshift of 0.92.
Only one object, 7C~1820+6657, was 
undetected in our study; its faint magnitude is consistent with its 
provisional redshift of 2.98.

In summary, we now have photometry on all nine of the high redshift radio
galaxies with either no redshift information or uncertain redshifts in the 
7C-{\sc iii} sample of 54 objects. All six objects with uncertain redshifts
have near-infrared magnitudes consistent with their tentative spectroscopic 
redshifts. Of the three objects with no redshift information, two have 
photometry consistent with them having redshifts $\sim 1.5$, and the one 
remaining object (7C~1748+6703) is probably at $z\sim 3$. 

\begin{figure}
\plotone{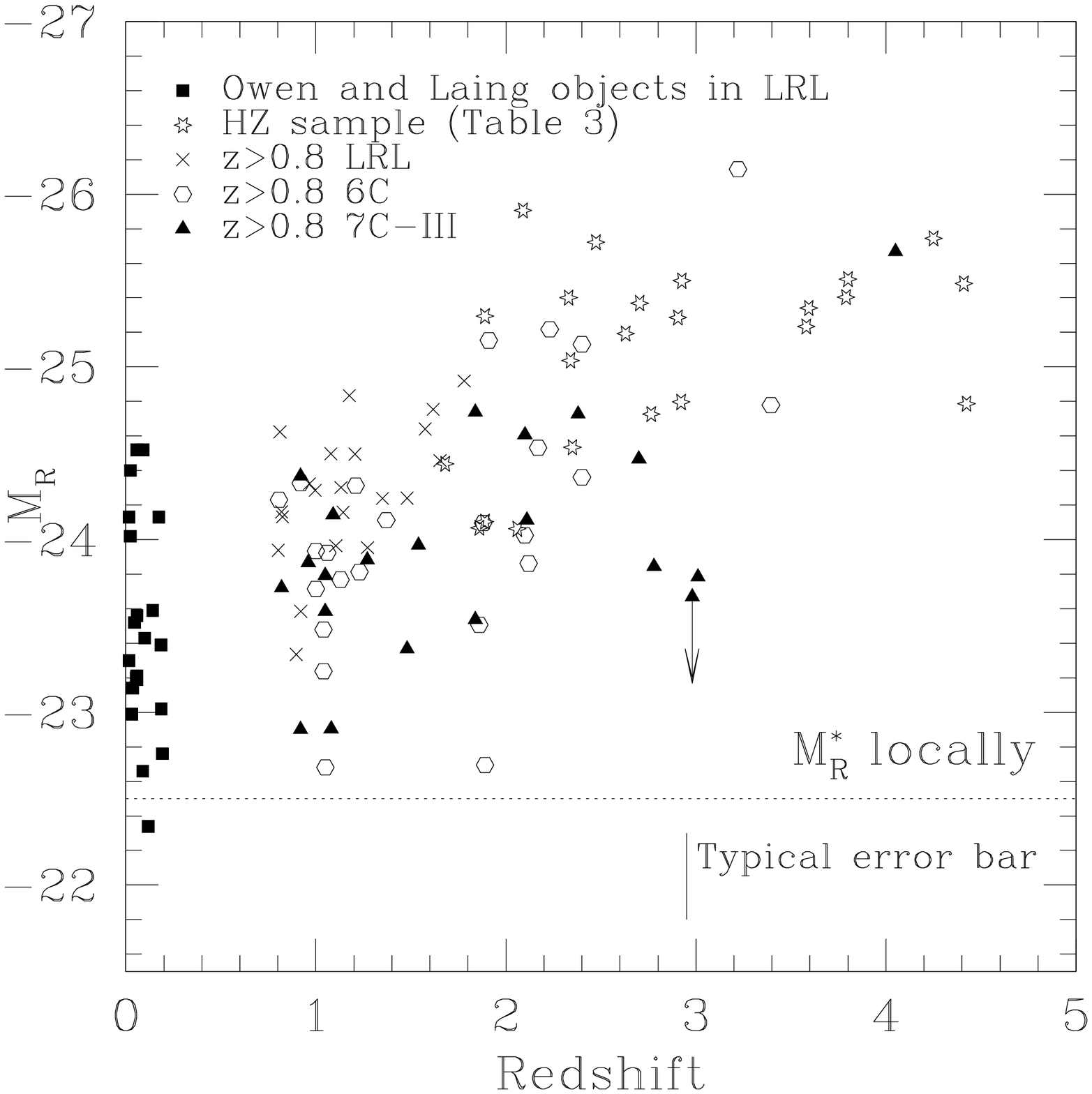}
%\figcaption[figure3.ps]{The rest-frame $R$-band absolute magnitudes. 
\caption{The rest-frame $R$-band absolute magnitudes. 
The value of $M_{R}^{*}$ locally is indicated by the dotted
line. A typical error bar for the 7C objects is indicated; the other 
samples generally have smaller error bars. \label{fig3}}
\end{figure}

\subsection{Emission line contamination of broad-band magnitudes}

Line contamination of continuum magnitudes 
is expected to be small in nearly all the 7C objects, as the correlation of 
emission line and radio luminosities means that, on average, 
emission line fluxes will be much less than those from objects in
radio-brighter samples. Whereas the $z\stackrel{>}{_{\sim}}2$ objects studied 
by Eales \& Rawlings (1996) typically have line contaminations of a few tens 
of percent, the 7C objects, which are selected at radio fluxes about four to 
five times lower, should have line contaminations 
$\stackrel{<}{_{\sim}} 10$ \%. We have listed in Table 2
the estimated percentage contribution to the observed flux for all 
objects with bright emission lines in the observed band. These 
are based on the spectroscopy of Lacy et al.\ (1999b), using the median 
Ly$\alpha/$H$\alpha$+[N{\sc ii}] ratio of 0.7 in Eales \& Rawlings (1996),
the line ratios in the composite radio galaxy spectrum of McCarthy (1993) and 
the H$\alpha$/H$\beta$ ratios of Koski (1978) to  
estimate the strengths of lines falling in the observed near-infrared. As 
expected, most objects have emission line contributions of $<10$ \% . For 
only one object, 7C 1802+6456, is the emission line contamination expected
to be $>$20 \% .
%This gives [O{\sc ii}]/$H$\alpha$+[N{\sc ii}] = 0.4.

\subsection{Estimation of half-light radii}

The seeing 
during most of our IRTF observations was very good, and it was clear from 
looking at the images that there was a wide range in the scale sizes of the 
hosts. We therefore selected a subsample consisting of the 15 sources with 
grade $\alpha$ or $\beta$ redshifts in the range 
$0.8<z<2.7$ in Lacy et al.\ (1999b) 
for which we had IRTF images. Although not complete, this 
sample should be representative of the 7C objects in this redshift range.
To estimate the scale sizes we convolved model elliptical (de Vaucouleurs) 
galaxies having a range in half-light radius $r_{\rm hl}$ from 
0.1 to 6.7 arcsec and zero ellipticity with a PSF from a nearby star
in each IRTF image. The model was subtracted from the data and the minimum 
in the sum of the squares of the residuals found, along with the approximate 
$\pm 1 \sigma$ range. An example of this process is shown in Fig.\ 2. 
The results were also checked by eye to see that 
the fits were reasonable. Nearby companion objects and discrete sub-components
were subtracted prior to the fits. Although this is far from a full host 
galaxy model (in particular we made no attempt to discriminate between disc 
and de Vaucoulours fits), we feel that it should
give a fair estimate of the scale size, and was all that the signal:noise in 
the images typically justified. 

To obtain a major axis scale size to compare to other work, we need to 
assume a mean ellipticity, $e$, and correct our scale sizes with this. Roche 
et al.\ and Govoni et al.\ (2000) derive very similar mean values of 
$e \approx 0.2$ for the 6C radio galaxies at $z\sim 1$ and nearby radio 
galaxy hosts respectively, so this value has been assumed to correct the mean 
scale sizes quoted in Section 4 by multiplying the scale size by the square 
root of the axial ratio corresponding to $e=0.2$, namely $1.06$. 

For two objects we have been able to compare the scale sizes with those
measured on images taken with the Hubble Space Telescope (HST). 
7C~1758+6719 ($z=2.70$) was observed with WFPC2 
through the F702W filter; details of the observations will be given in a
future paper (Lacy et al.\ in preparation). For this rest-frame UV
image an exponential disk with $r_{\rm hl} = 0.21$ 
arcsec was found to be a good fit to the radial profile. This can be 
compared with the estimate of $r_{\rm hl}$ from the IRTF image
of 0.35 arcsec with a $1 \sigma$ range of 0.12 -- 1.0 arcsec, 
which we consider fair agreement. 7C~1754+6420 $(z=1.09)$
was observed through the F675W filter (Ridgway \& Lacy in preparation). 
Again, the rest-frame UV emission was better fit by a disk than a de 
Vaucouleurs profile. In this case the scale size came out significantly smaller
than the estimate from the IRTF image, 0.7 arcsec in the HST image compared 
to a best fit of 4.5 arcsec and a range of 1.6 -- 13 arcsec for the 
IRTF image. This was despite the subtraction of a nearby companion to the 
south which was successfully removed from both the HST and IRTF images.
The cause of this discrepancy is not clear. This image had the worst seeing 
of any of our IRTF images (1.1 arcsec), but the scale size measured in the 
IRTF image is much larger than the seeing HWHM. The radial profile of 
the IRTF image is, however, not at all well fit by the disk model from 
the HST image. Therefore perhaps the most likely 
cause of this discrepancy is that the galaxy has a UV-bright disk component 
embedded in a much larger scale-size elliptical.

\subsection{Other data from the literature}

To increase the number of high redshift objects in our study we have added 
the results of photometry of $z\sim 2-5$ radio 
galaxies by van Breugel et al.\ (1998). We also estimated scale sizes from 
their figure 2 for objects which appeared to be dynamically-relaxed 
ellipticals (seven out of the eight of their radio galaxies with $z<3$).
We applied $k$- and aperture corrections in a consistent manner
to that for the 7C-{\sc iii} radio sources. We have also added those
$1.6<z<2.4$ objects without a strong point source contribution
from the NICMOS study of radio 
galaxies by Pentericci (1999; and Pentericci et al.\ 2000), 
including the estimates of scale sizes 
for the five out of the nine objects in this redshift range for 
which Pentericci considers it possible to fit meaningful scale sizes. 
We also include the low luminosity $z=4.42$ radio galaxy VLA~123642+621331 
discovered in the flanks of the Hubble Deep Field (Waddington et al.\ 1999)
and 53W002 at $z=2.239$ (Windhorst, Mathis \& Keel 1992).

The resulting high redshift (HZ) sample is detailed in 
Table 3. It mostly contains 
objects of radio luminosity comparable to the 3C radio galaxies
at $z\sim 1$. As these objects (particularly those at $z>3$) were generally
observed outside of rest-frame $R$-band, and are mostly not from complete
samples, the results including the HZ sample will not be as reliable, though 
as we shall see the same trends seem to exist with or without this sample. 

\begin{deluxetable}{llclclrc}
\footnotesize
\tablecaption{The ``HZ'' sample of  other 
high redshift radio galaxies from the literature
\label{tbl-3}}
\tablewidth{0pt}
\tablehead{
\colhead{Object} & \colhead{$z$} &\colhead{$\phi /^{''}$} &
\colhead{Mag.\ } & 
\colhead{$M_{R}$} & \colhead{$r_{\rm hl}/^{''}$} & \colhead{\% lines} 
& \colhead{ref.}}
% A: (1,0,50) B:(0.3,0.7,50), MR is in 63.9 kpc ap.

\startdata
6C 0140+326& 4.41 & 8 &$K_s=20.0$ & -25.5 &    & 10&1\\
MRC 0152-209&1.89 & 4 &$H=18.7$   & -25.3 &0.2 &  6&2\\
MRC 0156-252&2.09 & 4 &$H=18.4$   & -25.9 &    &  5&2\\
USS 0211-122&2.34 & 4 &$H=19.6$   & -25.0 &0.73&  3&2\\
MRC 0324-228&1.89 & 4 &$H=19.9$   & -24.1 &    &  0&2\\
4C 60.07   & 3.79 & 8 &$K^{'}=19.3$&-25.4 &    &  0&1\\
4C 41.17   & 3.80 & 8 &$K_s=19.2$ & -25.5 &    &  0&1\\
B3 0744+464& 2.926& 8 &$K_s=18.5$ & -25.5 &0.92&  0&1\\
MRC 0943-242&2.922& 8 &$K=19.2$   & -24.8 &0.70&  0&1\\
MG 1019+0534&2.765& 8 &$K=19.1$   & -24.7 &0.97&  0&1\\
3C 257     & 2.474& 8 &$K=17.8$   & -25.7 &1.55& 35&1\\
VLA 123642+621331&4.424&1.5&$K=21.4$&-24.8&0.29& 0&3\\
4C 1243+036& 3.581& 8 &$K=19.3$   & -25.2 &    &  0&1\\
USS 1410-001&2.33 & 4 &$H=19.3$   & -25.4 &    &  9&2\\
8C 1435+635& 4.251& 8 &$K=19.5$   & -25.7 &    &  0&1\\
MRC 1707+105& 2.35& 4 &$H=18.3$   & -24.5 &1.62&  0&2\\
53W002      & 2.239&- &$K=19.2$\tablenotemark{a}&-24.1  &1.1 & 20&4\\
MRC 2025-218&2.630& 8 &$K^{'}=18.5$&-25.2 &0.64&  0&1\\
MRC 2048-272&2.06 & 4 &$H=20.2$   &-24.1  &0.64&  0&2\\
MG 2121+1839&1.861& 8 &$K=18.7$   &-24.1  &3.07&  0&1\\
MG 2144+1929&3.594& 8 &$K^{'}=19.2$& -25.3&    &  0&1\\
TX 2202+128 &2.704& 8 &$K=18.4$   &-25.4  &0.19&  0&1\\
MRC 2224-273&1.68 & 4 &$H=19.2$   &-24.4  &0.2 & 28&2\\ 
4C28.58     &2.905& 8 &$K=18.7$   &-25.3  &    &  0&1\\
\enddata

\tablenotetext{a}{Total magnitude}

\tablecomments{References: (1) van Breugel et al.\ (1998); (2) Pentericci 
(1999); (3) Waddington et al.\ (1999); (4) Windhorst et al.\ (1992)}

\end{deluxetable}

\section{Discussion}

\subsection{Radio-luminosity and redshift dependence of absolute magnitudes}

In Fig.\ 3 we plot the rest-frame $R$-band magnitudes against redshift
for several samples of radio galaxies. We plot 
the $z>0.8$ 7C-{\sc iii} objects with spectroscopic redshifts and infrared 
imaging, $z>0.8$ 6C radio galaxies with photometry from Eales et al.\ (1997), 
$z>0.8$ 3C radio galaxies in the Laing, Riley \& Longair (1983; LRL) complete 
sample with photometry from Best, Longair \& R\"{o}ttgering (1998) 
[apart from 3C22 which is a lightly-reddened quasar (Rawlings et al.\ 1995) 
and therefore excluded from the sample, 3C~175.1 ($z=0.92$) which has 
photometry from Ridgway \& Stockton (1997) and 3C~263.1 ($z=0.824$) which 
has photometry from Eales (personal communication)]. We have also added 
local radio galaxies in LRL from Owen \& Laing (1989) and the HZ sample of 
galaxies of Table 3. All these objects, with the exception of three FRIs out
of the 24 objects in the Owen \& Laing LRL sample are either FRII 
or compact steep-spectrum sources. All the $z>0.8$ 7C-{\sc iii} sources 
at $z>0.8$ are well above the FRI/FRII boundary in radio luminosity, with 
$L_{R (151)} \stackrel{>}{_{\sim}} 10^{26} {\rm WHz^{-1}sr^{-1}}$ compared 
to the FRI/FRII boundary at $L_{R (151)} 
\sim 10^{25} {\rm WHz^{-1}sr^{-1}}$.

There is a clear trend for redshift and absolute magnitude to correlate, 
even if the incomplete HZ sample is 
excluded. The 3C, 6C and 7C samples are all complete samples, selected on 
the basis of low frequency radio flux only, and are nearly completely
identified. Thus the only selection effect which needs to be considered for 
these samples is the tendency for redshift and luminosity to correlate within 
each flux limited sample. With the wide range in radio luminosities in the 
complete samples at $z\sim 1$, however, we can separate out the luminosity 
dependence. This is illustrated in Fig.\ 4, where we have plotted absolute 
magnitude against radio luminosity for 3C, 6C and 7C galaxies in the redshift 
range $0.8<z<1.4$. In this redshift range the mean magnitude in 3C is 
$M_{R} = -24.17 \pm 0.13$, whereas in 6C it is $-23.79 \pm 0.14$ and in 
7C-{\sc iii} $-23.78 \pm 0.15$. This suggests that only the most 
radio-luminous objects have slightly brighter (by $\approx 0.4$ mag) hosts
[see also Rawlings et al.\ (1998) where preliminary $K$-band photometry on the 
7C-{\sc i} and 7C-{\sc ii} samples is presented]. 

We can think of two 
possible explanations for this radio luminosity dependence 
of host magnitude at $z\sim 1$. The first is a contribution
from AGN-related light, e.g.\ emission lines and reddened and/or scattered 
quasar light (Eales \& Rawlings 1996). Rigler et al.\ (1992), however, 
show that the fraction of emission from $z\sim 1$ 3C radio galaxies
which is aligned with the radio jet axis, and therefore probably closely
related to the AGN activity, contributes only 
$\approx 10$\% of the light in the observed $K$-band. Also, Simpson, 
Rawlings \& Lacy (1999) argue that reddened quasar light is responsible for 
$\stackrel{<}{_{\sim}} 10$\% of the observed near-infrared emission from 3C 
radio galaxies on the basis of 3$\mu$m imaging. Thus we believe that 
only $\sim 10-20$ per cent of the host luminosity can be accounted for 
by the AGN, not the $\approx 40$ per cent required to explain the higher
host luminosities of the 3C galaxies.
The second possibility is that the radio luminosity may be correlated with
the mass of the host. This could arise, for example, if 
correlation of black hole 
mass and the mass of the spheroidal component of galaxies which is
claimed to be present at 
low redshift (e.g.\ Magorrian et al.\ 1998) is appearing
at the highest radio luminosities (Roche et al.\ 1998). We have 
argued (Lacy, Ridgway \& Trentham 2000; see also Willott et al.\ 1999) 
that most radio galaxies are sub-Eddington accretors, but at the 
highest luminosities corresponding to the highest black hole masses, even the 
radio galaxies may be accreting at near Eddington rates. Thus at these
highest luminosities we might expect a host galaxy mass -- AGN luminosity 
(and therefore host luminosity) correlation to appear. Also, the mass
of the host may correlate with the density of the intergalactic medium 
confining the radio source, which would enhance the radio luminosity. 

\begin{figure}
\plotone{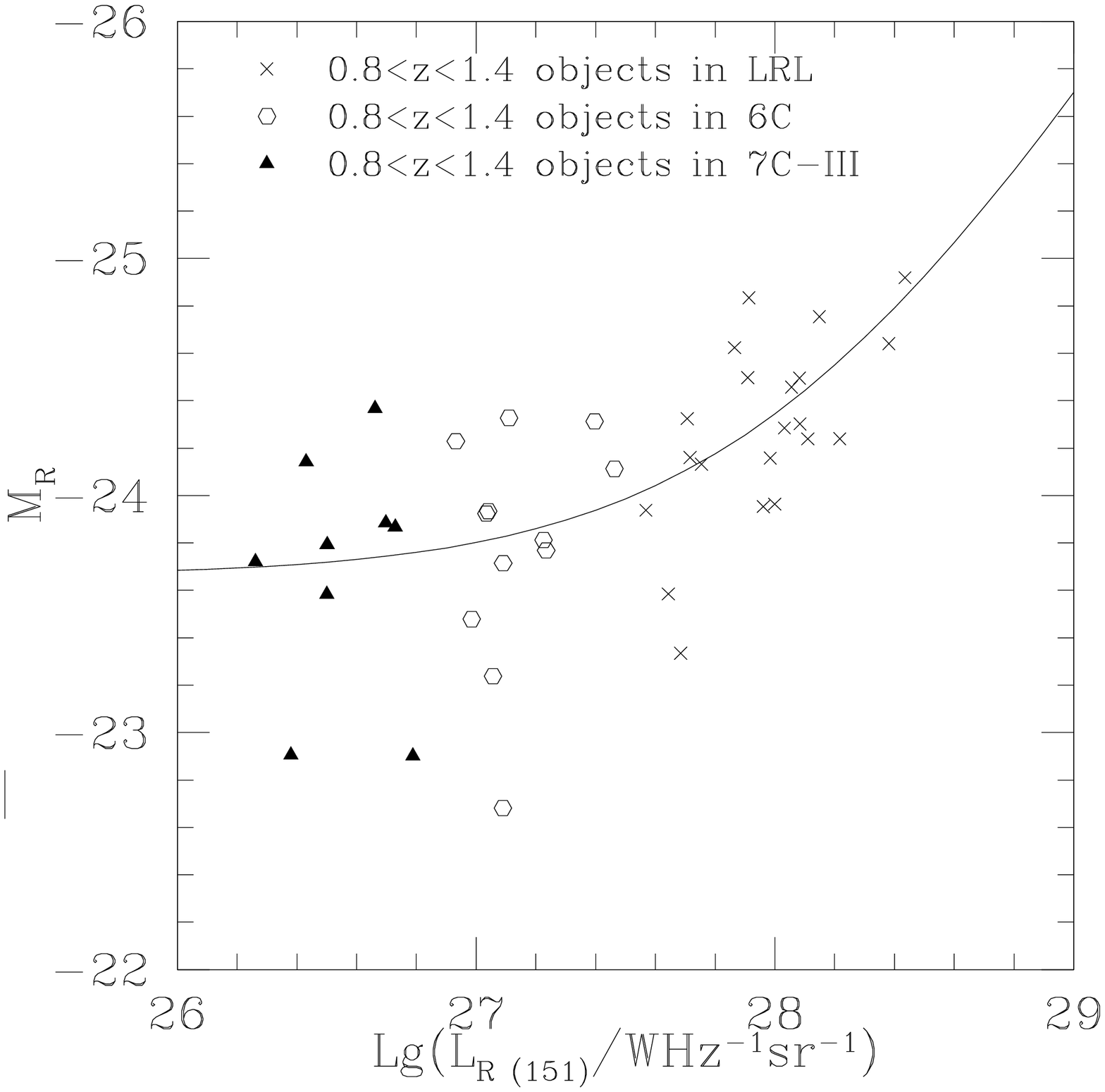}
%\figcaption[figure4.ps]{The luminosity dependence of host galaxy 
%magnitude at $z\sim 1$. The line represents the model of equation (1).}
\caption{The luminosity dependence of host galaxy 
magnitude at $z\sim 1$. The line represents the model of equation (1).}
\end{figure}

We have made a crude model of the luminosity dependence 
by assuming a contribution from AGN-related light to a base host optical 
luminosity $L_0$. We model this contribution as a combination of 
a component due to emission lines and another component due to other
AGN-related light. (We separate the emission line contribution, as the 
equivalent width of the emission lines is a strong function of redshift, 
unlike the other AGN-related emission which we expect to be much less 
redshift dependent.) Both these components should be proportional to
the AGN luminosity, which we assume to be proportional to $L_R (151)$ to the 
power 0.8 (Serjeant et al.\ 1998; Willott et al.\ 1999), i.e. 
\begin{equation}
L_{\rm host}(z) = L_0(z) + (\alpha + \beta (z)) L_{R (151)}^{0.8}, 
\end{equation}
where $\alpha L_{R (151)}^{0.8}$ is the contribution of AGN-related light 
other 
than emission lines and $\beta (z) L_{R (151)}^{0.8}$ is the contribution 
due to 
emission lines. The base host luminosity at $z\approx 1$ was set to 
$L_0(1) = 10^{23} {\rm WHz^{-1}}$, or
$M_{R}\approx -23.7$, close to the mean of the 7C and 6C data, and 
$(\alpha + \beta(1))$ fit by eye to be $\approx 3.5 {\rm W^{0.2}Hz^{-0.2}}$ 
(Fig.\ 4). The emission
line contribution to the total near-infrared luminosities of the 
3C radio galaxies at these redshifts is $\sim 10$\% (Rawlings et al.\ 1997; 
Rawlings Eales \& Lacy 1990), or about 20\% of the overall 
luminosity-dependent correction. We have therefore set 
$\alpha=2.8 \, {\rm W^{0.2}Hz^{-0.2}}$ and 
$\beta (1) = 0.7 \, {\rm W^{0.2}Hz^{-0.2}}$. 

Using this formula we can then correct the host magnitudes for AGN-related 
light and produce plot of $L_0$ versus redshift. [Rather than 
attempt to model the redshift dependence of $\beta$ 
we have used emission line contamination estimates from 
Eales \& Rawlings (1997) and van Breugel et al.\ (1998) to explicitly correct
the magnitudes in the 6C and high redshift samples (where these were unknown
the object was omitted from the plot and correlation analysis), and our own 
estimates to correct the 7C-{\sc iii} magnitudes.] The results are shown in 
Fig.\ 5, which shows that there is indeed a correlation of $L_0$ with redshift.
This was confirmed using the Kendall Tau 
statistic generalized to include limits, as implemented in 
the {\sc iraf.stsdas}  program {\sc bhkmethod} (Isobe \& Feigelson 1990), 
which returned a $\tau = 0.52$ for the 106 objects from the 3C, 6C, 7C and 
HZ samples after correction. This corresponds to a probability 
that there is {\em no} correlation between $L_0$ and redshift of $<0.0001$. 
However, this probability increases to 0.13 if the HZ sample is removed.
We have also conducted the same statistical test before correction 
for radio luminosity-dependent effects in equation (1). This gave a 
$\tau = 0.95$ and a probability of no correlation of $<0.0001$, removing the 
HZ sample reduced this to $\tau=0.66$, again with the probability of no 
correlation of $<0.0001$.

\begin{figure}
\plotone{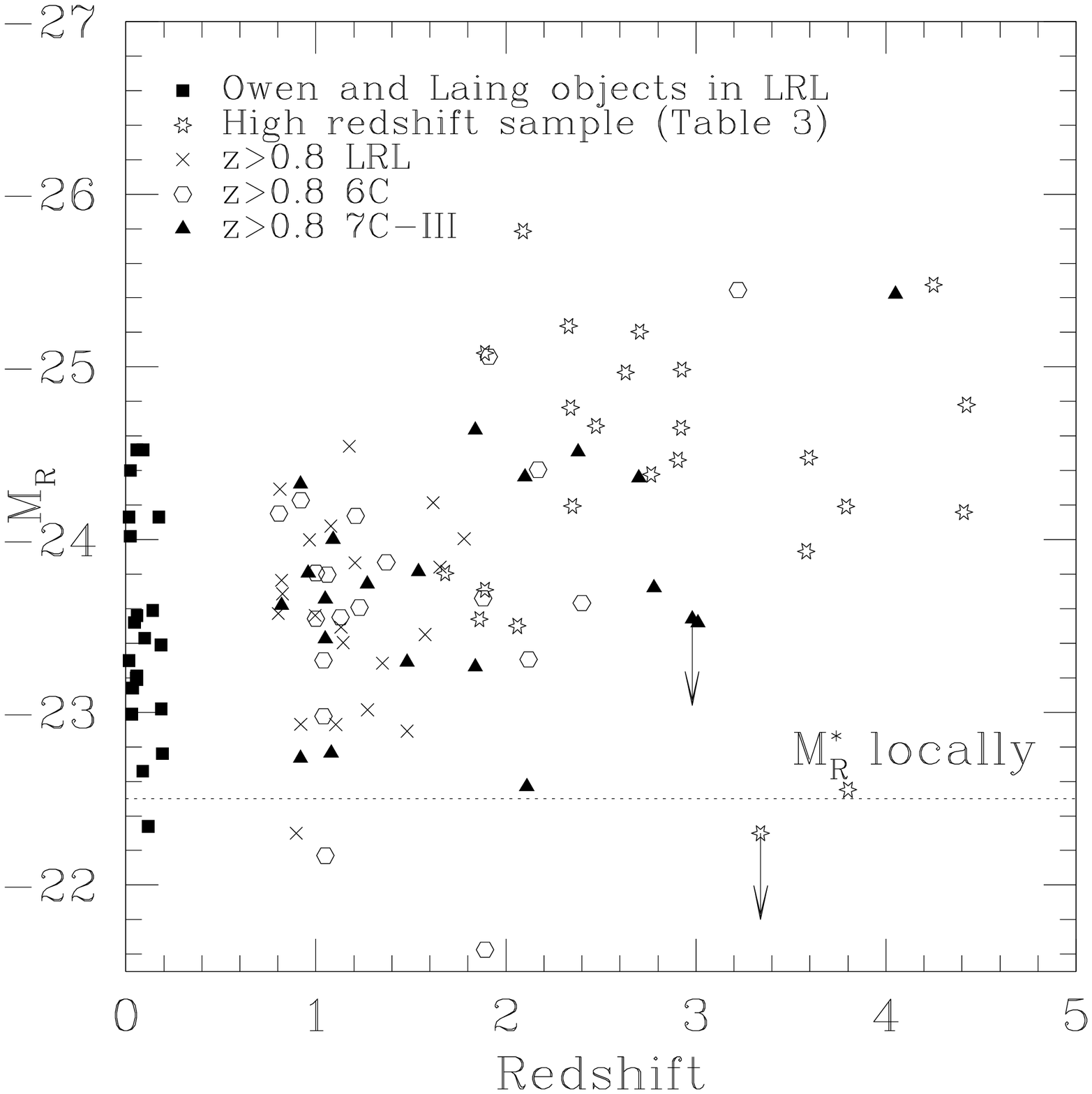}
%\figcaption[figure.ps]{The redshift dependence of radio galaxy 
\caption{The redshift dependence of radio galaxy 
absolute magnitude after approximate correction for enhancements which 
correlate with radio luminosity, including emission lines.}
\end{figure}

\subsection{The dispersion in the host luminosities}

The low scatter in the magnitudes of radio galaxies in the $K-z$ Hubble 
Diagram out to $z\stackrel{>}{_{\sim}}3$ has long been used as an argument for
a high redshift of radio galaxy formation (e.g.\ Lilly 1989). This low scatter 
is thought to come about because radio galaxies form at $z>3$ then  
evolve along similar passive evolution tracks 
to end at radio galaxy hosts today. With a few modifications this simple
picture still seems to be basically valid. Fig.\ 3 and Table 4 show that 
the dispersion in absolute magnitudes to $z\sim 4$ is indeed very low.
After correction 
for AGN-contamination, however, the dispersion in $L_0$ for the  $z>1.8$
objects seems to be higher than for the uncorrected objects (Table 4). 
As discussed above, the correlation of absolute magnitude with redshift
is also less tight after correction. 

At first sight it seems paradoxical that correction
for luminosity-dependent effects should {\em increase} the scatter in 
the absolute magnitudes. However, this can be be understood if the 
luminosity-dependent contributions $(\alpha + \beta)L_{R (151)}^{0.8}$ 
are approximately equal to the highest values of $L_0$. (The $z>1.8$ 
objects have a very narrow range in $L_R$ as the range in flux of these 
objects is low, all except VLA 123642+621331 
having 151 MHz fluxes between 0.5 and 10 Jy.)
The addition of 
the luminosity-dependent contributions can then boost the objects with
low $L_0$ by a significant factor, whereas the addition to the objects 
with high $L_0$ results in only a relatively small fractional increase
in the luminosity (see also de Vries 1999).  
The result of this is a reduction of the scatter in the total luminosities. 

We have used the F-test for variances to examine the statistical significance
of the increase in scatter of the $L_0$ values. Comparing the objects with
$0.8<z<1.8$ to those with $1.8<z<2.8$, we find that the increase in the 
sample standard deviation
from $\sigma_{n-1}=0.53$ to $\sigma_{n-1}=0.93$ is significant at the 0.1 \% 
level. At $z>2.8$, our sample contains two upper limits on the magnitudes, so
we give a lower limit to the scatter of  $\sigma_{n-1}>0.97$. This is 
again significantly higher than that in the range $0.8<z<1.8$ (at 
the 0.2 \% level). 
Thus there is good evidence that the scatter in $L_0$ is increasing
with redshift, suggesting that we are close to the formation epoch of at 
least a significant fraction of the radio galaxy population. 

\onecolumn
\begin{table}
%\begin{center}
\caption{Scatter in radio galaxy absolute magnitudes as a function of redshift
\label{tbl-4}}
\begin{tabular}{ccrrrr}
\tableline\tableline
Redshift & $n$\tablenotemark{a} &\multicolumn{2}{c}{Mean $M_{R}$} & 
\multicolumn{2}{c}{$\sigma_{n-1}$\tablenotemark{b}}\\
 range   && uncorrected & corrected & uncorrected & corrected\\
\tableline
$0.8\leq z<1.8$ &44&-24.0 & -23.6 & 0.5 & 0.5 \\
$1.8\leq z<2.8$ &25&-24.6 & -24.2 & 0.7 & 0.9 \\
$z\geq 2.8$     &15&$>$-24.3& $>$-25.1&$>$0.7 & $>$1.0\\
\tableline
\end{tabular}

\tablenotetext{a}{Number of objects in each sub-sample}

\tablenotetext{b}{Standard deviation of the sample.}

\tablecomments{$M_{R}$ is measured in a 63.9 kpc metric aperture. Uncorrected
magnitudes have $k$-corrections only, no corrections for emission line or AGN 
luminosity have been applied. Corrected magnitudes have, in addition 
to $k$-corrections, 
emission-line contamination corrections and a correction for AGN luminosity
as detailed in Section 4.1.}
%\end{center}
\end{table}
\twocolumn

\begin{figure}
\plotone{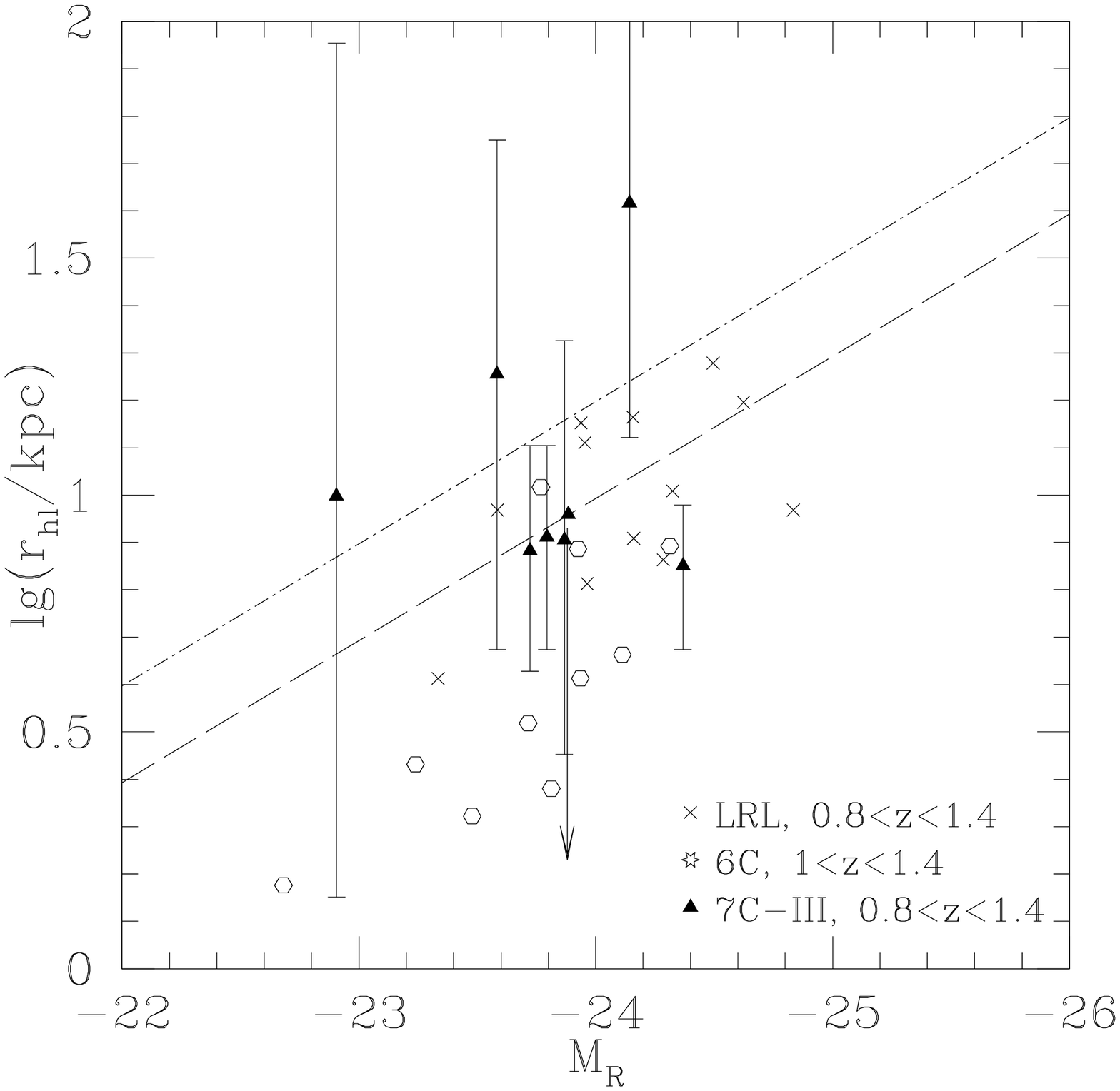}
%\figcaption[figure6.ps]
\caption{
The dependence of half-light radius on host absolute magnitudes. Errors are 
shown for the 7C objects only. The dot-dashed line represents the 
correlation observed for local ellipticals (equation (2) with 
$\Delta R = 0$); the dashed line the 
same relationship corrected to $z\sim 1$ using the mean properties of the
3C hosts (i.e.\ with $\Delta R = 0.68$).}
\end{figure}

\begin{figure}
\plotone{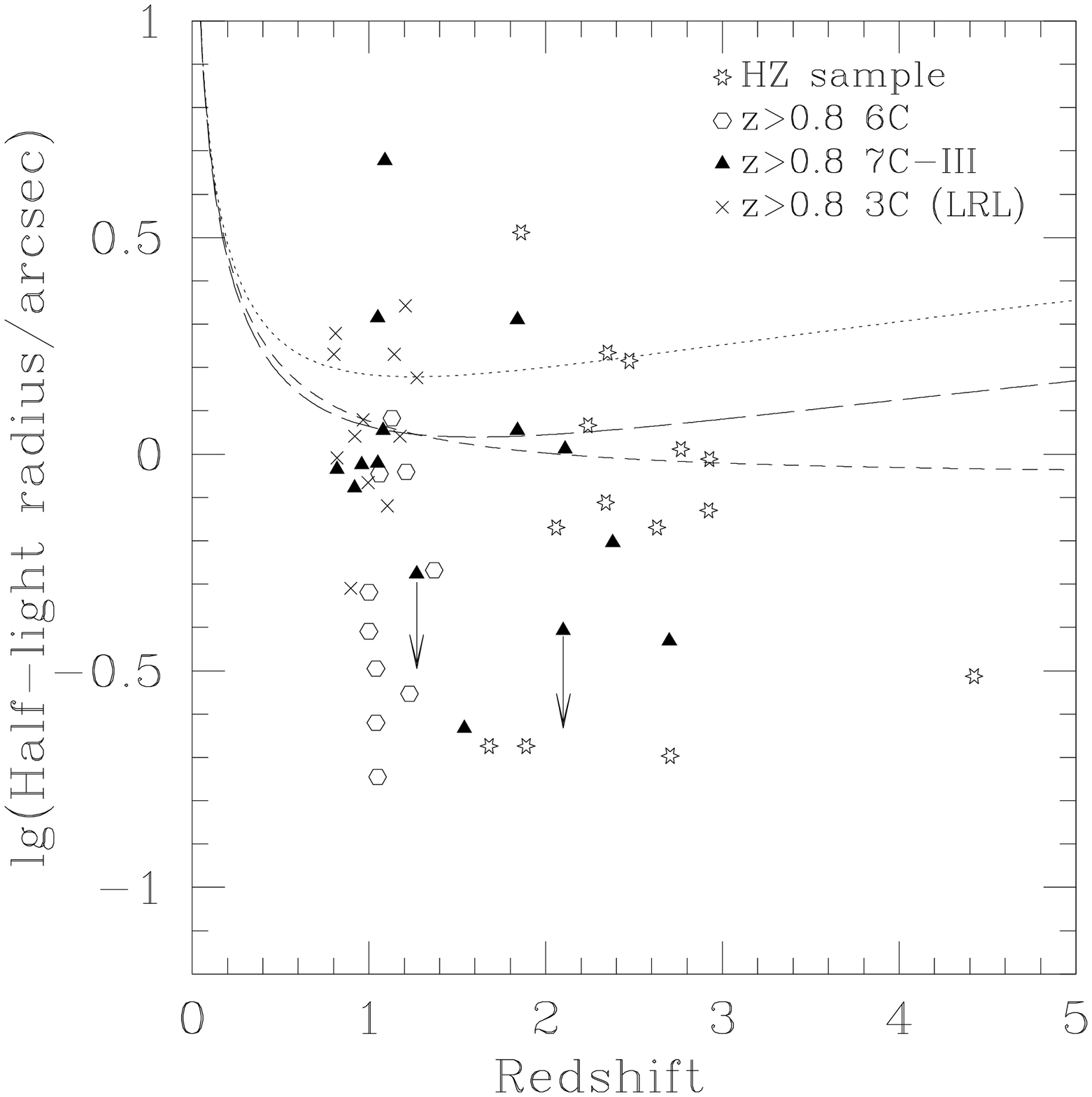}
%\figcaption[figure7.ps]{Half-light radius in arcseconds versus 
\caption{Half-light radius in arcseconds versus 
redshift. The lines show the angular size of a standard rod of length
equal to the mean of the half-light radii of local FRII host galaxies (13 kpc)
as a function of redshift in an $\Omega_{M}=1, \Omega_{\Lambda}=0$ cosmology
(dotted line), an $\Omega_{M}=0.3, \Omega_{\Lambda}=0.7$ cosmology 
(long-dashed line)
and an $\Omega_{M}=0, \Omega_{\Lambda}=0$ cosmology (short-dashed line).}
\end{figure}

\subsection{The scale sizes of the hosts}

Roche et al.\ (1998) have shown that the scale sizes of the 
$1<z<1.4$ 6C galaxies are significantly smaller than their 
more radio luminous 3C counterparts as measured by Best et al.\ (1997), 
suggesting a strong dependence of scale size on AGN luminosity. 
In this section we re-examine this using our data on the 7C objects, and
further data on 3C scale sizes published recently. 

To compare with the 6C results, 
we have defined complete samples of 7C-{\sc iii} and 3C radio galaxies from
LRL in the redshift range $0.8<z<1.4$ (we took the lower limit of 0.8 
so as to include more 3C and 7C objects; only two 6C objects lie in the
redshift range $0.8<z<1$, and we do not feel their exclusion will 
significantly affect the results). For the
3C objects we have used size estimates from the 2D modelling of McLure \& 
Dunlop (2000) or the Keck $K$-band images of Ridgway \& Stockton (1997) where 
available, and otherwise from Best et al.\ (1997). No estimates were available
for three of the sixteen objects in the 3C sample, and we have omitted 3C356
due to confusion over the true identification. One of the objects missing
from the 3C sample were omitted from the sample of Best et al.\ (1997)
(3C~263.1), and the other two, 3C~13 and 3C~368 have $K$-band 
structures which are clearly affected by AGN-related emission as they are 
closely aligned with the radio axes. 
%These estimates of scale sizes seem to be 
%mostly fairly consistent with those measured in HST NICMOS images and plotted 
%in Zirm et al.\ (2000). 
Of the 7C-{\sc iii} objects, only 
one, 7C~1816+6710, has no scale size estimate as its image was taken in 
poor seeing, although the scale size of 7C~1742+6346 is an upper limit 
only. All the 7C scale sizes were increased by a factor of 1.06 to correct
for an expected mean ellipticity of 0.2.

In Fig.\ 6 we plot the scale sizes of the 3C, 6C and 7C radio galaxies in the 
samples defined above against host absolute magnitude. Although the scatter is 
large it does seem that the 7C-{\sc iii} radio galaxies have scale sizes 
consistent with their magnitudes when compared to the 3C radio galaxies, 
although the 6C objects are plotting below the general trend. The 
mean scale size for the 3C sample is $11.1 \pm 1.4$ kpc and that for the 
6C sample $4.7\pm 0.9$kpc. For the 7C-{\sc iii} sample, we find a 
median scale size of $0.9\pm 0.4$ arcsec, which, when corrected for 
a mean ellipticity of 0.2, corresponds to about $8\pm 3$ kpc.
(Using the whole subset of 7C-{\sc iii} sources in the range $0.8<z<2.7$
with measured scale sizes we obtained the same result.) The mean for the 
7C-{\sc iii} is thus between
that of the 6C and 3C samples, and  statistically consistent with both. 

A further check is provided by the effective-radius -- galaxy magnitude 
relationship discussed by Roche et al.\ (1998), 
\begin{equation}
 {\rm lg} (r_{hl}/{\rm kpc}) = -0.3 (M_R + \Delta R + 20.01), 
\end{equation}
where $\Delta R$ is the amount by which the hosts have brightened by passive 
evolution from $z=0$. To set $\Delta R$, we used the mean scale size of the 
3C galaxies in our $0.8<z<1.4$ sample of 11.1 kpc, and the 
mean magnitudes of the same objects using the photometry of Best et al.\ 
(1998). This gave $\Delta R=0.68$. We have plotted the relationship of 
equation (2) on Fig.\ 6, for both $\Delta R=0$ and $\Delta R =0.68$.
The 7C radio galaxies are consistent with this relation for 
$\Delta R =0.68$ (within the scatter), 
but most of the 6C radio galaxies plot significantly below the line. 

We therefore believe that a combination 
of poor seeing in both the Roche et al.\ (1998) and Best et al.\ (1998) 
studies, which lead to poor scale size estimates, combined
with small number statistics in the Roche et al.\ study, 
can probably explain the result
of Roche et al.\ without the requirement for scale-size to depend very
strongly on radio luminosity, though clearly better images of both the 6C 
and 7C objects will be needed to be certain of this.

We have also investigated possible redshift dependences of scale size with the
addition of data on objects in the HZ sample listed in Table 3. 
Fig.\ 7 shows the results of plotting the scale size in arcsec against 
redshift. The mean half-light radius for local hosts of classical double 
FRIIs from Owen \& Laing (1989) and Govoni et al.\ (2000), 
$\approx 13\pm 2.5$ kpc, is also plotted as a line in this figure for 
three different cosmologies.
There is some evidence for a weak trend for scale size to decrease
with increasing redshift, but with a lot of scatter. In an attempt to reduce
the scatter we have tried using a correction based on equation (1), by 
introducing a correction in the log of the scale size:
\[ \Delta {\rm lg} r_{\rm hl} = -0.3 (M - M_0(z)), \]
where $M$ is the absolute $R$-band magnitude after correction for line 
contamination, and $M_0 (z)$ is $L_0 (z)$ converted to absolute magnitude. 
With this correction, using the 7C, 6C, 3C ($0.8<z<1.4$) and HZ 
samples there is a weak anticorrelation of scale
size with redshift, significant at about the 5\% level using the {\sc 
bhkmethod}.    

Inspection of Fig.\ 7, and the comparison to low redshift FRII hosts, 
shows that the scatter in scale sizes is large at all redshifts, and suggests
that any evolution in scale size is weak. FRI hosts at low redshift, 
however, have significantly brighter hosts and larger scale sizes at a given 
radio luminosity (e.g.\ Owen \& Ledlow 1994), suggesting that merging in the 
richer 
cluster environments associated with these objects may be significant if they 
are the descendents of more luminous sources at high redshift. An argument 
against this is provided by a recent study of brightest cluster galaxies
(BCGs) by Burke, Collins \& Mann (2000), who show that, when selection effects
are properly taken into account, the BCGs of the most X-ray luminous clusters
brighten with redshift (to a similar extent as the FRII hosts, in fact), 
and thus there is little evidence of significant merger activity for 
redshifts $0<z<1$. However, because the FRII hosts are significantly
less luminous than BCGs, mergers that would enhance the luminosity
of a BCG by only a few percent would enhance that of an FRII host by a 
much larger factor. Nevertheless, the lack of evidence for merger activity 
in even rich clusters suggests that FRII hosts, which are generally found in 
poor cluster environments (e.g.\ Wold et al.\ 2000), are probably not 
significantly affected by merging, at least at $z<1$.  
%Indeed, using the scale sizes of radio galaxy hosts as standard 
%rods may even be feasible, although the large intrinsic range in scale sizes 
%and the difficulty of measuring them accurately means that large numbers of 
%deep near-infrared images will be required. So far we see no sign of the 
%upturn in scale sizes that one might expect in a Universe with near-critical 
%density (Fig.\ 7). If we really can neglect merging,
%the data are clearly more consistent with a low $\Omega_{M}$,
%but more data are required before we can begin to 
%constrain $\Omega_{M}$ with any degree of confidence. 

\begin{figure}
\plotone{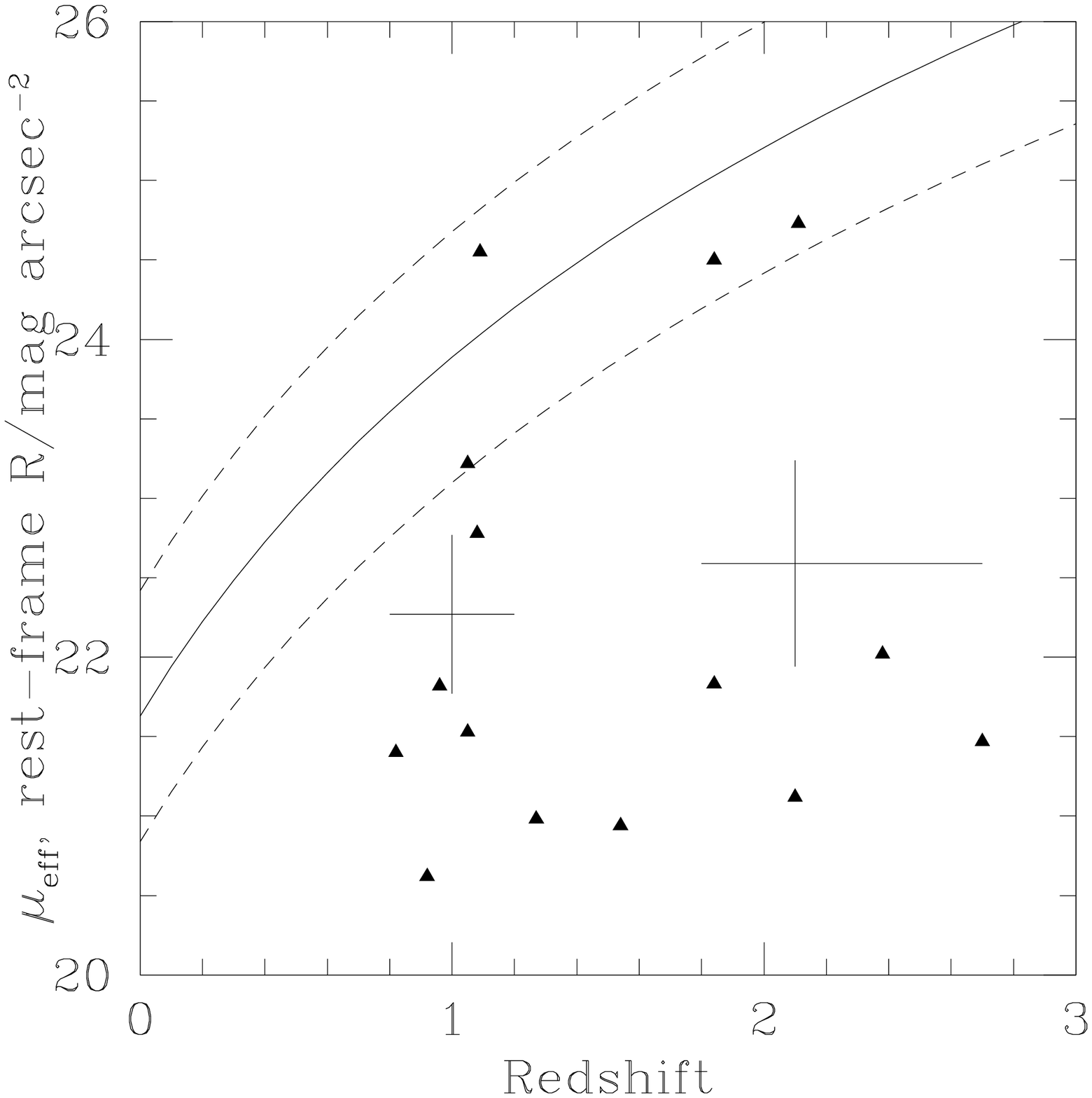}
%\figcaption[figure8.ps]{The effective surface brightness as a function of
\caption{The effective surface brightness as a function of
redshift for the $0.8<z<2.7$ 7C-{\sc iii} galaxies observed with the IRTF, 
after correction for emission line contamination.
The crosses are the mean values of the surface brightness in the 
redshift range indicated by the horizontal bars and with an error in the 
mean indicated by the vertical bar. The mean local value for FRII 
radio galaxies from Govoni et al.\ (2000)
and its expected behaviour with 
redshift is indicated by the solid line, with the dashed lines showing the 
$\pm 1 \sigma$ range. \label{fig8}}
\end{figure}

\subsection{Surface brightness evolution}

To demonstrate that the redshift dependence of host magnitude is 
independent of the assumed cosmology
we have plotted in Fig.\ 8 the effective surface brightness 
(defined here as the flux within $r_{\rm hl}$ divided by $\pi r_{\rm hl}^2$ 
converted into magnitudes) in the 
rest-frame $R$-band versus redshift for the sub-sample of 7C-{\sc iii} 
galaxies with estimated scale sizes. We have 
compared this to the cosmological surface brightness dimming expected 
in a standard expanding Universe with a Friedmann-Robertson-Walker metric. 
The surface brightness decreases with redshift much less 
rapdly than the  
prediction for a non-evolving stellar population (solid line in Fig.\ 8). 
This demonstrates that the brightening of the hosts with 
redshift is a real evolutionary effect, and not one produced by an incorrect
choice of  $\Omega_{M}$ and $\Omega_{\Lambda}$.

\section{The evolution of radio galaxy hosts}

The evolution of host magnitude and rest-frame surface brightness with 
redshift, the lack of strong evolution in scale sizes in the subset of 
objects with ``undisturbed'' morphologies, and the low scatter 
($< 1$ magnitude) in the absolute magnitudes up to at least $z\sim 3$
are all consistent with  an early  formation epoch ($z\stackrel{>}{_{\sim}}3$)
for most, and perhaps all, radio galaxy 
hosts. At first sight these results are in conflict with standard CDM-based
models for galaxy formation, in which hierarchical structure formation 
occurs in a ``bottom up'' fashion with the most massive galaxies forming late.
Radio galaxies at $z\sim 2$ are much brighter than the average quasar host
predicted by the models of Kauffmann \& Haehnelt (2000), for example. To 
compare our radio galaxy sample to these predictions we have assumed an
equivalent quasar luminosity equal to the mean of the $1<z<3$ 7C quasars, 
$M_{B} \approx -25$ (Willott et al.\ 1998). For quasars of this luminosity, 
the Kauffmann \& Haehnelt models predict a mean host magnitude of $M_V\approx 
-21.5$ compared to our radio galaxies with a luminosity-corrected median 
$M_R \approx -23.5$ (corresponding to $M_V \approx -23$). In contrast, the 
hosts of radio-quiet quasars at $z\sim 2$ seem to fit this model quite well 
(Ridgway et al.\ 2000). 

Between $z\sim 1.3$ and $z\sim 2.3$, there is evidence for an increasing 
scatter in base host luminosity. This suggests we are close to the 
formation epoch at $z\sim 2.3$. There are two other pieces of evidence which
point to strong evolution in the host galaxies and which appear at 
$z\stackrel{>}{_{\sim}} 2.5$. First, the
observations of Pentericci (1999; and Pentericci et al.\ 2000) and  
van Breugel et al.\ (1998) show that
the morphologies of radio galaxy hosts at $z<2.5$ are mostly relaxed 
ellipticals, whereas at higher redshifts ($z>2.5$ in the HST NICMOS $H$-band 
images of Pentericci et al.\ 
and $z>3$ in the ground-based $K$-band images of van Breugel 
et al.) hosts with clumpy structures aligned with the radio source axis 
become much more common\footnote{van Breugel et al.\ argue 
that this is not simply due to the
$z>3$ objects being observed at shorter rest-frame wavelengths, as 
lower redshift objects observed at the same rest-frame wavelengths as the 
$z>3$ objects still look like relaxed ellipticals.}. Second, 
the detection rate of continuum submillimeter emission from hot dust seems to 
rise rapidly at $z>2.5$ (Archibald et al.\ 2000). The 6C and 7C 
samples are completely identified with luminous host galaxies out to at 
least $z\approx 3$, however, which would suggest the bulk of the stars had 
formed and/or merged onto the host by $z\sim 3$.
One way around a high formation redshift for all objects would be to assume 
a systematic delay between a host 
forming the bulk of its stars and the switching on of powerful radio 
jets (for example, if the black hole needs time to accrete enough mass). 
Radio galaxies would then always be seen with luminous hosts. However, 
the strong evolution in the hosts seen by van Breugel et al.\ around 
$z\sim 3$ suggests that their nature is still changing up to at least that 
epoch.

Taken together, the   
evidence suggests that the epoch of radio galaxy host formation was
essentially over by $z\sim 3$, in the sense that most objects had formed
the bulk of their stars, although further star-formation and 
merger activity continued in many objects up to $z\sim 2$. 
As radio galaxies can be found out to $z>5$, however (van Breugel et al.\ 
1999), and there is no sign of a substantial drop in the space density of 
radio galaxies out to $z>4$ relative to their peak space density at 
$z\sim 2.5$ (Jarvis et al.\ 1999), this formation epoch must last at 
least a Gyr. In contrast, field ellipticals with evolved stellar 
populations and luminosities $\sim L_*$ seem to become rarer at $z>1$ 
(Zepf 1998; Barger et al.\ 1999; 
Dickinson 1999), suggesting a later formation epoch, perhaps $z\sim 1-2$.

A recent variation on the hierarchical 
models, the so-called Anti-Hierarchical Baryonic Collapse Model 
(Granato et al.\ 2000) may be able to explain these observations.
This predicts that the massive hosts of powerful AGN may have formed at 
$z\sim 3$. This model uses the same CDM assumptions
as the standard scenarios, but assumes that the dense gas in the most massive
haloes can collapse and form stars in a massive spheroidal component more 
rapidly than that in the smaller haloes associated with normal galaxies.
Star formation is switched off in these objects by winds from the AGN and 
the starburst at high redshift. 
If radio galaxy hosts are indeed typical of 
the most massive elliptical galaxies and their progenitors this model should 
be applicable to them, and would explain their anomalous evolution relative 
to generally less luminous field ellipticals.

\acknowledgments

We thank Wim de Vries, Michael Gregg and Wil van Breugel helpful discussions, 
the referee for a useful report, and the staff at the IRTF and 
Lick Observatory for their assistance. 
We are also very grateful to Chris Willott for providing the photometric 
redshift estimates for those 7C-{\sc iii} radio galaxies without 
spectroscopic redshifts prior to publication.
The IRTF is operated by the University of Hawaii on behalf of NASA.
AJB acknowledges support from the Cambridge Institute of Astronomy PPARC 
observational rolling grant, ref.~no.~PPA/G/O/1997/00793, and a NICMOS 
postdoctoral fellowship while at Berkeley (grant NAG\,5-3043). 
This work was performed under the auspices of the U.S. Department of
Energy by University of California Lawrence Livermore National
Laboratory under contract No. W-7405-Eng-48, and was partly
based on observations with the NASA/ESA Hubble Space Telescope, obtained at 
the Space Telescope Science Institute, which is operated by the Association 
of Universities  for Research in Astronomy, Inc. under NASA contract 
No. NAS5-26555.


\begin{thebibliography}{}
\bibitem[Archibald et al.(2000)]{Arc00} Archibald E.N., Dunlop J.S., Hughes
D.H., Rawlings S., Eales S.A., Ivison R.J., 2000, MNRAS, submitted
(astro-ph/0002083)
\bibitem[Barger et al.(1999)]{Bar99} Barger A.J., Cowie L.L., Trentham N., 
Fulton E., Hu E.M., Songaila A., Hall D., AJ, 117, 102
\bibitem[Best et al.(1997)]{BLR} Best P.N., Longair M.S., 
R\"{o}ttgering H.J.A., 1997, MNRAS, 292, 758
\bibitem[Burke et al.(2000)]{Bur00} Burke D.J., Collins C.A., Mann R.G., 
2000, ApJL, in press (astro-ph/0002185)
\bibitem[de Vries 1999]{WimTh} de Vries W., 1999, PhD thesis, University of 
Groningen
\bibitem[Dickinson (1999)]{Dic99} Dickinson M., 1999, in, Hammer F., et al., 
eds, Building Galaxies: From the Primordial Universe to the Present, 
Proceedings of the XIXth Moriond Astrophysics Meeting. Ed.\ Fronti\`eres, 
p.\ 257, in press
%\bibitem[Dunlop \& Peacock (1993)]{DP93} Dunlop J.S., Peacock J.A., 1993, 
%MNRAS, 263, 936
\bibitem[Eales \& Rawlings (1993)]{ER93} Eales S.A., Rawlings S., 1993, ApJ, 
411, 67
\bibitem[Eales \& Rawlings (1996)]{ER96} Eales S.A., Rawlings S., 1996, ApJ, 
460, 68
\bibitem[Eales et al.(1997)]{Ea97} Eales S., Rawlings S., Law-Green D., Cotter
G., Lacy M., 1997, MNRAS, 291, 593
\bibitem[Govoni et al.(2000)]{Gov00} Govoni F., Falomo R., Fasano G., 
Scarpa R., 1999, A\&A, in press (astro-ph/9910469)
\bibitem[Granato et al. 2000]{Gra00} Granato G.L., Silva L., Monaco P., 
Panuzzo P., Salucci P., De Zotti G., Danese L., 2000, MNRAS, in press 
(astro-ph/9911304)
%\bibitem[Ivison et al.(1998)]{Ivi98} Ivison R.J., Dunlop J.S., Hughes D.H., 
%Archibald E.N., Stevens J.A., Holland W.S., Robson E.I., Eales S.A.,
%Rawlings S., Dey A., Gear W.K., 1998, ApJ, 494, 211
\bibitem[Fioc \& Rocca-Volmerange]{PegI} Fioc M., Rocca-Volmerange B., 1997, 
A\&A, 326, 950
\bibitem[Isobe \& Feigelson (1990)]{IF90} Isobe T., Feigelson E.D., 1990, BAAS,
22, 917
\bibitem[Jarvis et al.\ (2000)]{Jar00} Jarvis M.J., Rawlings S., Willott C.J., 
Blundell K.M., Eales S., Lacy M., 2000, in, van Breugel W.J., Bunker,A.J., 
eds, The Hy-Redshift Universe. ASP, in press 
\bibitem[Kauffmann \& Haehnault 2000]{KH00} Kauffmann G., Haehnault M., 2000, 
MNRAS, 311, 576
\bibitem[Koski (1978)]{K78} Koski A.T., 1978, ApJ, 223, 56
%\bibitem[Lacy et al. (1999a)]{align} Lacy M., Ridgway S.E., Wold M., 
%Lilje P.B., Rawlings S., 1999a, MNRAS, 307, 420
\bibitem[Lacy et al. (1999a)]{PIII} Lacy M., Kaiser, M.E., Hill G.J., 
Rawlings S., Leyshon G., 1999a, MNRAS, 308, 1087
\bibitem[Lacy et al. (1999b)]{PIV} Lacy M., Rawlings S., Hill G.J., 
Bunker A.J., Ridgway S.E., Stern D., 1999b, MNRAS, 308, 1096
\bibitem[Lacy, Ridgway \& Trentham (2000)]{MBHs} Lacy M., Ridgway S.E., 
Trentham N., 2000, in, Biretta et al., eds,
Lifecycles of Radio Sources, New Astronomy Reviews, in press
\bibitem[Laing, Riley \& Longair (1983)]{LRL} Laing R.A., Riley J.M., Longair
M.S., 1983, MNRAS, 204, 151 (LRL)
\bibitem[Lilly (1989)]{Lil89} Lilly S.J., 1989, ApJ, 340, 77
\bibitem[Magorrian et al.(1998)]{Mag98} Magorrian J., Tremaine S., 
Richstone D., Bender R., Bower G., Dressler A., Faber S.M., Gebhardt K., 
Green R., Grillmair C., Kormendy J., Lauer T., 1998, AJ, 115, 2285
\bibitem[McCarthy (1993)]{Mc93} McCarthy P.J., 1993, ARA\&A, 31, 639
\bibitem[McLean et al.(1993)]{Gem1} McLean I.S., et al., 1993, Proc.\ SPIE, 
1946, 512 
\bibitem[McLean et al.(1994)]{Gem2} McLean I.S., et al., 1994, Proc.\ SPIE, 
2198, 457
\bibitem[McLure \& Dunlop (2000)]{McL3C} McLure R.J., Dunlop J.S., 2000, 
MNRAS, submitted (astro-ph/9908214) 
\bibitem[McLure et al. (1999)]{McL99} McLure R.J., Kukula M.J., Dunlop J.S., 
Baum S.A., Hughes D.H., 1999, MNRAS, 308, 377
\bibitem[Owen \& Laing (1989)]{OL89} Owen F.N., Laing R.A., 1989, MNRAS, 
238, 357
\bibitem[Owen \& Ledlow (1994)]{OL94} Owen F.N., Ledlow M.J., 1994, in, 
Bicknell G.V., Dopita M.A., Quinn P.J., eds, The First Stromolo Symposium: 
The Physics of Active Galaxies. ASP Conf.\ Ser.\ Vol.\ 54, p.\ 319 
\bibitem[Pentericci (1999)]{PentTh} Pentericci L., 1999, PhD Thesis, 
University of Leiden
\bibitem[Pentericci et al.(2000)]{Pen00} Pentericci L., McCarthy P.J., 
R\"{o}ttgering H.J.A., Miley G.K., van Breugel W.J.M., Fosbury R., 2000, 
ApJ, submitted
\bibitem[Rawlings et al.(1990)]{REL90} Rawlings S., Eales S., Lacy M., 1990,
MNRAS, 251, 17P
\bibitem[Rawlings et al.(1995)]{Raw95} Rawlings S., Lacy M., Sivia D.S., 
Eales S.A., 1994, MNRAS, 274, 428
\bibitem[Rawlings et al.(1997)]{Raw97} Rawlings S., Blundell K.M., Lacy M., 
Willott C.J., Eales S., 1998, in, Bremer M.N., Jackson N., 
P\'{e}rez-Fournon I., eds, Cosmology with the New Radio Surveys. Kluwer, 
Dordrecht, p.\ 171
\bibitem[Ridgway \& Stockton (1997)]{RS97} Ridgway S.E., Stockton A.N., 
1997, AJ, 114, 511
\bibitem[Ridgway et al. 2000]{R99} Ridgway S.E., Heckman T., Calzetti D., 
Lehnert M.,  2000, in, Biretta et al., eds, Lifecycles of Radio Galaxies, 
New Astronomy Reviews, in press
\bibitem[Rigler et al.(1992)]{Rig92} Rigler M.A., Lilly S.J., Stockton A.N., 
Hammer F., Le F\`{e}vre O., 1992, ApJ, 385, 61
\bibitem[Roche, Eales \& Rawlings (1998)]{RER98} Roche N., Eales S., Rawlings 
S., 1998, MNRAS, 297, 405
\bibitem[Serjeant et al.(1998)]{Ser98} Serjeant S.B.G., Rawlings S., 
Lacy M., Maddox S.J., Baker J.C., Clements D., Lilje P.B., 1998, MNRAS, 
294, 494 
\bibitem[Simpson, Rawlings \& Lacy (1999)]{ThIm} Simpson C., Rawlings S., 
Lacy M., 1999, MNRAS, 306, 828
\bibitem[Stevens (1999)]{Ste99} Stevens R., 1999, DPhil.\ Thesis, University 
of Oxford
\bibitem[van Breugel et al.(1998)]{vB98} van Breugel W.J.M., Stanford S.A., 
Spinrad H., Stern D., Graham J.R., 1998, ApJ, 502, 614
\bibitem[van Breugel et al.(1999)]{vB99} van Breugel W., De Breuck C., 
Stanford S.A., Stern D., R\"{o}ttgering H., Miley G., 1999, ApJ, 518, L61
\bibitem[Waddington et al.(1999)]{Wad99} Waddington I., Windhorst R.A., 
Cohen S.H., Partridge R.B., Spinrad H., Stern D., 1999, ApJ, 526, L77
\bibitem[Willott et al. (1998)]{7CQSO} Willott C.J., Rawlings S., 
Blundell K.M., Lacy M., 1998, MNRAS, 300, 625 
\bibitem[Willott et al. (1999)]{Wemmlns} Willott C.J., Rawlings S., 
Blundell K.M., Lacy M., 1999, MNRAS, 309, 1017
\bibitem[Willott et al. (2000)]{Wrlf} Willott C.J., Rawlings S., 
Blundell K.M., Lacy M., 2000, in preparation
\bibitem[Willott, Rawlings \& Jarvis (2000)]{WRJ00} Willott C.J., Rawlings S., 
Jarvis M.J., 2000, MNRAS in press (astro-ph/9910422)
\bibitem[Windhorst et al.(1992)]{WMK92} Windhorst R., Mathis D.F., Keel W.C., 
1992, ApJ, 400, L1
\bibitem[Wold et al.(2000)]{Wol00} 
Wold M., Lacy M., Lilje P.B., Serjeant S.B.G., 
2000, MNRAS, submitted (astro-ph/9912070)
\bibitem[Zepf (1997)]{Zep97} Zepf S.E., 1997, Nat, 390, 377
\end{thebibliography}
\end{document}